\documentclass[twocolumn,showpacs,preprintnumbers,amsmath,amssymb,aps,prl,showkeys,final]{revtex4} %two column final form 
\usepackage{graphicx}% Include figure files
\usepackage{dcolumn}% Align table columns on decimal point
\usepackage{bm}% bold math

\usepackage{calc}
\usepackage{epsfig}
\input{pstricks}\input{pst-node}
\usepackage{chngpage}
\newcommand{\setPageForLLNLCover}[2]{%
\newlength{\textwidthOld}%
\setlength{\textwidthOld}{\textwidth}%
\newlength{\textheightOld}%
\setlength{\textheightOld}{\textheight}%
\newlength{\topmarginOld}%
\setlength{\topmarginOld}{\topmargin}%
\newlength{\textwidthNew}%
\setlength{\textwidthNew}{6.5in}%
\newlength{\textheightNew}%
\setlength{\textheightNew}{9.5in}%
\newlength{\oddsidemarginNew}%
\newlength{\topmarginNew}%
\setlength{\oddsidemarginNew}{\oddsidemargin}
\setlength{\topmarginNew}{\topmargin-0.3cm}
\newlength{\oddsidemarginOld}%
\setlength{\oddsidemarginOld}{\oddsidemargin}%
\changepage{\textheightNew-\textheightOld}{\textwidthNew-\textwidthOld}{\oddsidemarginNew-\oddsidemarginOld}{\oddsidemarginNew-\oddsidemarginOld}{}{\topmarginNew-\topmarginOld}{}{}{}%
}%
\newcommand{\resetPageFromLLNLCover}{%
\changepage{-\textheightNew+\textheightOld}{-\textwidthNew+\textwidthOld}{-\oddsidemarginNew+\oddsidemarginOld}{-\oddsidemarginNew+\oddsidemarginOld}{}{-\topmarginNew+\topmarginOld}{}{}{}%
}%
\newcommand{\makeLLNLCover}[7]{%
\setPageForLLNLCover{#6}{#7}%
\thispagestyle{empty}% no number of this page
\newcommand{\logoWidth}{1.65in}% 
\psset{xunit=1.cm,yunit=1.cm,runit=1.cm}%
\begin{pspicture}(0,0)(17,24.)
\rput(2.3,11.5){}
\rput(11.2,23.){\parbox{12.0cm}{\large\bf%
\begin{flushright}   %%PREPRINT NUMBER: give as UCRL-JRNL-XXX XXXX
Preprint \\
#1
\end{flushright}
}}
\rput(10.5,18){\parbox{12.0cm}{%\sffamily\bfseries\Huge\noindent%
\fontsize{24.88}{30pt}\usefont{OT1}{cmss}{bx}{n}
\begin{flushleft}
#2
\end{flushleft}
}}
\rput(10.5,13.){\parbox{12.0cm}{%\sffamily\LARGE\noindent%
\fontsize{17.28}{18pt}\usefont{OT1}{cmss}{m}{sl}
\begin{flushleft}
#3
\end{flushleft}
}}
\rput(10.5,9.5){\parbox{12.0cm}{% \sffamily\large\noindent%
\fontsize{14}{16pt}\usefont{OT1}{cmss}{m}{n}
This article was submitted to #4}}
\rput(10.5,7.5){\parbox{12.0cm}{% \sffamily\bfseries\LARGE\noindent%
\fontsize{20.74}{22pt}\usefont{OT1}{cmss}{bx}{n}
\begin{flushleft}
#5
\end{flushleft}
}}
\rput(10.5,-1.){\parbox{12.0cm}{%
Approved for public release; further dissemination unlimited}}
\end{pspicture}
\clearpage 
\changetext{.625in}{}{}{}{}
\thispagestyle{empty}% no number of this page
\vglue5\baselineskip
\begin{center}
{\bf DISCLAIMER}
\end{center}
\noindent
This document was prepared as an account of work sponsored by an agency of the United
States Government.  Neither the United States Government nor the University of California
nor any of their employees, makes any warranty, express or implied, or assumes any legal
liability or responsibility for the accuracy, completeness, or usefulness of any
information, apparatus, product, or process disclosed, or represents that its use would
not infringe privately owned rights. Reference herein to any specific commercial product,
process, or service by trade name, trademark, manufacturer, or otherwise, does not
necessarily constitute or imply its endorsement, recommendation, or favoring by the
United States Government or the University of California.  The views and opinions of
authors expressed herein do not necessarily state or reflect those of the United States
Government or the University of California, and shall not be used for advertising or
product endorsement purposes.

This is a preprint of a paper intended for publication in a journal or proceedings. Since
changes may be made before publication, this preprint is made available with the
understanding that it will not be cited or reproduced without the permission of the
author.
\vskip2\baselineskip
This research was supported under the auspices of the U.S. Department of Energy by
the University of California, Lawrence Livermore National Laboratory under
contract No.  W-7405-Eng-48.
\vfill
\begin{center}
Approved for public release; further dissemination unlimited
\end{center}
\clearpage
\changetext{-.625in}{}{}{}{}
\resetPageFromLLNLCover
\setcounter{page}{1}
}
\usepackage{program}
\newcommand{\reynolds}{\mbox{Re}}
\newcommand{\mach}{\mbox{Ma}}
\newcommand{\machBending}{\mbox{Mb}}

\newcommand{\schmidt}{\mbox{Sc}}
\newcommand{\schmidtGamma}{\mbox{Sg}}

\newcommand{\ubar}{\bar{u}}
\newcommand{\vbar}{\bar{v}}

\newcommand{\Gammabar}{\bar{\Gamma}}

\newcommand{\uvbar}{\bar{\uv}}

\newcommand{\jump}[1]{[\![ #1 ]\!]}

\newcommand{\rstrain}{\bf{D}}
\newcommand{\Kstar}{K^{*}}

\newcommand{\cross}{\times}
\newcommand{\transp}{T}

\newcommand{\derx}{\partial_x}
\newcommand{\dery}{\partial_y}

\renewcommand{\vec}[1]{{\mathbf #1}}

\newcommand{\grad}{\vec{\nabla}}
\renewcommand{\div}{\vec{\nabla}\cdot}
\newcommand{\laplace}{\nabla^2}

\newcommand{\pder}[2]{\frac{\partial #1}{\partial #2}}
\newcommand{\matder}[1]{\frac{D#1}{Dt}}

\newcommand{\eps}{\varepsilon}

\renewcommand{\vec}[1]{{\mathbf #1}}

\newcommand{\xivec}{\vec{\xi}}

\newcommand{\normalvec}{\nv}
\newcommand{\tangvec}{\sv}

\newcommand{\jv}{\mathbf{ j}}

\newcommand{\nv}{\mathbf{ n}}

\newcommand{\qv}{\mathbf{ q}}

\newcommand{\sv}{\mathbf{ s}}

\newcommand{\uv}{\mathbf{ u}}

\newcommand{\Dv}{\mathbf{ D}}

\newcommand{\soapSection}[1]{}
\newcommand{\soapSubsection}[1]{}
\newcommand{\soapSubsubsection}[1]{}

\newcommand{\etal}{{\em et al.}}

\newcommand{\Ord}{{O}}

\newcommand{\ucrlNumber}{UCRL-JRNL-214578}

\newcommand{\dateVersion}{October 24, 2005}
\newcommand{\paperTitle}{A viscous compressible model of soap film flow 
       and its  
       equivalence with the Navier-Stokes equations 
       }
\begin{document}

%%------------------------------------------BEGIN LLNL COVER PAGE----------
\normalbaroutside %%%NEEDED for program.sty
%%\renewcommand{\baselinestretch}{1.}
%\singleSpacing
\makeLLNLCover{\ucrlNumber}{\paperTitle}
    {Petri Fast \\
     Center for Applied Scientific Computing\\
     Lawrence Livermore National Laboratory}
 {Phys. Fluids}{\dateVersion}{0in}{0in}
%%\renewcommand{\baselinestretch}{1.7}
%\doubleSpacing
%%------------------------------------------END LLNL COVER PAGE-------------
\setcounter{page}{0}
\thispagestyle{empty}

\preprint{\ucrlNumber}

\title{\paperTitle }

\author{Petri Fast}
\email{fast1@llnl.gov}
% \altaffiliation[Also at ]{Physics Department, XYZ University.}
\affiliation{Center for Applied Scientific Computing,
Lawrence Livermore National Laboratory, 
Livermore, CA}

\date{\dateVersion}

\begin{abstract}
We present a quasi-two dimensional model of flowing
soap films that bears striking similarity to the 
compressible Navier-Stokes equations.
The variation in soap film thickness that is commonly used
for flow visualization in experiments is analogous
to density variations in the Navier-Stokes equations.
When the soap film flow velocity is comparable to the Marangoni elastic 
wave velocity we recover the
compressible Navier-Stokes equations and
the soap film behaves like a two-dimensional isothermal viscous gas.
\end{abstract}


% PACS, the Physics and Astronomy Classification Scheme.
%% 47.10.+g 	General theory (Fluid Dynamics)
%% 47.40.-x 	Compressible flows; shock and detonation phenomena
%% 02.30.Mv 	Approximations and expansions
%%

%%% IN Vorobieff, PRL v. 81, p. 1417 (1998)
% 92.60.Ek 	Convection, turbulence, and diffusion
% 47.27.Gs 	Isotropic turbulence; homogeneous turbulence
% 67.40.Vs 	Vortices and turbulence

%                             % Classification Scheme.

\maketitle

%\section{\label{sec:level1}First-level heading:\protect\\ The line
%break was forced \lowercase{via} \textbackslash\textbackslash}
%\subsection{\label{sec:level2}Second-level heading: Formatting}

\soapSection{Introduction}
%\input{pfsec-intro.tex}
%
%%..MOTIVATION
%
Fast flowing soap films have been used extensively 
as an experimental realization
of two-dimensional fluid dynamics\cite{rutgersReview}. 
The purpose of this Letter is to present a new
quasi-two dimensional model of flowing
soap films that bears striking similarity to the 
compressible Navier-Stokes equations.
In a particular case, 
we can formally identify the 
viscous soap film model with the two-dimensional compressible 
Navier-Stokes  equations:
The variation in soap film thickness that is commonly used
for flow visualization in experiments is analogous
to density variations in the  Navier-Stokes equations in a 
quasi-two dimensional setting. The effective viscosity 
in our model is variable and depends on the film thickness.

The original papers~\cite{couderSoapFilm,gharib} 
lay out the experimental and theoretical foundation for
considering flow in flat soap films as classical two-dimensional flow.
A new experimental technique developed in the late 1990's\cite{rutgersReview}
further popularized this approach by
enabling very large, long-time stable, gravity driven soap films
that have been used for studies of two-dimensional 
turbulence~\cite{vorobieffRiveraEcke,batchelor2d,kraichnan2d},
fluid-structure interactions~\cite{zhangFilament} 
and shockwave dynamics~\cite{wenSoapShock2}.

Most previous theoretical work on modeling 
flowing variable thickness soap films has
considered them as represented by the two-dimensional constant density
incompressible Navier-Stokes equations\cite{couderSoapFilm}.
However, Vorobieff~\etal\cite{vorobieffRiveraEcke} have pointed out
in a series of papers significant thickness variations in
rapid soap film flows. 
Chomaz~\cite{chomaz2001} derived recently a new model of soap film
flow and argued that: (1) The model in an inviscid limit 
is equivalent to the  two-dimensional compressible
Euler equations, and (2) no correspondance 
with the compressible Navier-Stokes equations could be found.

In this Letter, we establish a connection between
the compressible Navier-Stokes equations and a 
model of flowing soap films.
We derive a quasi-two dimensional  viscous
compressible model of fast flowing soap film
that is applicable in subsonic and supersonic regimes. 
The model is derived systematically from the 
three-dimensional incompressible Navier-Stokes equations
coupled with a model of surfactant transport and capillary forces on the 
free surfaces.

\soapSection{A compressible model of soap film flow}

We consider a flowing soap film of thickness $H$ and width $L$
and scale $z\sim H$ and the lateral directions $x,y\sim L$.
In typical experiments $H\approx 10\mu$m and $L\approx 10$cm. We develop 
a thin film model using the small parameter $\eps=H/L$.
The film is driven by  an incoming flow rate $Q$ which determines a 
characteristic flow velocity $U=Q/HL$.
An important characteristic of soap films is the
Marangoni stress that arises from variations of the surface tension $\sigma$
that depend on the 
local surfactant concentration $\Gamma$ on the free surface $z=h(x,y,t)$.
We assume a linear surface tension model\cite{deWit}
$\sigma = \sigma_a - \sigma_r\Gamma$ with parameters $\sigma_a$, $\sigma_r$.
We scale the velocity $u,v\sim U$, $w\sim \eps U$,
and the pressure  $p\sim \eps\sigma_m/L$ to balance the capillary forces,
and define the mean surface tension 
$\sigma_m = \sigma_a - \sigma_r\Gamma_m$, and the 
mean surfactant concentration $\Gamma_m$.

The scaling leads to five nondimensional groups:
the Reynolds number $\reynolds = \rho UL/\mu_0$,
the elastic {\em (Marangoni)} Mach number
$\mach  = U(\rho H/(\sigma_r \Gamma_m))^{1/2}$,
the bending (capillary) Mach number
$  \machBending     = \eps^{-1}U(\rho H/\sigma_m)^{1/2}$,
the surface Schmidt number $\schmidtGamma = \mu_0/(\rho D_{surf})$ 
for the diffusion of $\Gamma$,
and the bulk Schmidt number $\schmidt = \mu_0/(\rho D_c)$ for 
the diffusion of $c$.   We define
relaxation constants  $\lambda, \Kstar$ that describe 
the exchange of surfactant concentration $\Gamma$ at the surface with
the surfactant concentration $c$ in the interstitial fluid. 

A new formulation of two-dimensional viscous compressible 
soap film flow is given by
\begin{eqnarray}
\label{eq:mom}
  \pder{(h\uv)}{t} &+& \div(h\uv\uv)
          = -\frac{1}{\mach^2}\grad\Gamma 
                + \frac{h}{\machBending^2}\grad\grad^2 h \\
\notag
\mbox{} &+& \div\left\{\frac{h}{\reynolds}
                       \left[ (\grad\uv+\grad\uv^{\transp}) 
                         + 2(\div\uv){\mathbf \cal I} \right]
                  \right\} \\
\label{eq:thickness}
  \pder{h}{t} &+& \div\left(h\uv\right) = 0 \\
\label{eq:surfactant}
   \pder{\Gamma}{t} &+& \div\left(\Gamma\uv\right)
      = ~~-\lambda \left( \Gamma - c \right) 
            + \frac{1}{\schmidtGamma\reynolds}\laplace\Gamma  \\
\label{eq:soapConcentration}
\pder{(hc)}{t} &+& \div(hc\uv)
      = \lambda \,\Kstar\left( \Gamma - c \right)
       + \frac{h}{\schmidt\reynolds}\laplace c
\end{eqnarray}
based on the primitive variable equations introduced
by Chomaz\cite{chomaz2001}.
Here ${\mathbf \cal I}$ is the identity tensor.
A full derivation of 
Eqs.~\eqref{eq:mom}--\eqref{eq:soapConcentration} is provided below
after we discuss some of the properties of this model.

A key result of this Letter is the new formulation of the 
viscous stress tensor 
$\eta(\grad\uv+\grad\uv^{\transp}) + 2\eta''(\div\uv){\mathbf\cal I}$ 
in the soap film model:
This is formally identical to the viscous terms in the
compressible Navier-Stokes equations
with dynamic viscosity $\eta=2h/\reynolds$ and dilatational viscosity
$\eta''= h/\reynolds $.  %, Ref.~\cite{batchelor}.
Actually, the viscous terms in
Chomaz~\cite[Eqs. (3.14)]{chomaz2001}, and 
Ida \&\ Miksis~\cite[Eqs. (14)--(18)]{idaMiksis2},
which were written out in nonconservative form,
are identical to the viscous term in Eq.~\eqref{eq:mom}.
The benefit of rewriting the viscous terms in conservative form
is that it reveals the similarity with the compressible Navier-Stokes 
equations. 
Past work\cite[Sec. 4.2 on p. 404]{chomaz2001} 
expressed doubts that a soap film model can be compared
with the compressible Navier-Stokes 
equations  since it appeared that the soap film 
viscous stress could not  be expressed in a form such as Eq.~\eqref{eq:mom}.

In fact,
Eqs.~\eqref{eq:mom}--\eqref{eq:soapConcentration}  are 
formally
equivalent to the compressible Navier-Stokes equations
in flows where dispersive effects are insignificant 
($1/\machBending\rightarrow 0$) and
surfactants  are insoluble ($\lambda\rightarrow 0$).
The thickness $h$ and concentration $\Gamma$ in the soap film model
correspond to the density and pressure, respectively, in the 
Navier-Stokes equations with a special equation-of-state.
We consider two special cases.

First, we consider a finite Reynolds number version of the ``inviscid 
supersonic soap film'' limit
in Ref.~\cite[Sec. 4.2]{chomaz2001}.
The interstitial soap concentration $c\equiv C_0$ is assumed to be constant.
Eqs.~\eqref{eq:mom}--\eqref{eq:soapConcentration} simplify to
\begin{eqnarray}\label{eq:nsFirst}
\pder{(h\uv)}{t} &+&\div(h\uv\uv)
       = \frac{-1}{\mach^2}\grad\Gamma \\
  \mbox{} &+& \frac{1}{\reynolds}\div\left\{ \eta[(\grad\uv+\grad\uv^{\transp})
                                        +2(\div\uv){\mathbf \cal I}]\right\}
 \notag\\
 \pder{h}{t} &+& \div(h\uv) = 0\\
 \pder{\Gamma}{t} &+&\div(\Gamma\uv) =0,\label{eq:nsLast}
\end{eqnarray}
where the dynamic and dilatational viscosity is $\eta=h$. The 
ratio $\Gamma/h$ is conserved along streamlines in this limit\cite{chomaz2001}
which corresponds to the behavior of $p/\rho$ in isothermal flows.
Therefore, the soap film Eqs.~\eqref{eq:nsFirst}--\eqref{eq:nsLast}
 are identical to 
the compressible Navier-Stokes equations for isothermal fluids without 
viscous heating. In the inviscid limit $\reynolds\rightarrow\infty$ these
equations are identical to Eq.~(4.13) in Ref.~\cite{chomaz2001}.

Second, we consider incompressible flow in a variable thickness soap film.
Eqs.~\eqref{eq:nsFirst}--\eqref{eq:nsLast}  simplify to
\[
\pder{(h\uv)}{t} +\div(h\uv\uv)
       = -\grad\phi                 
  + \frac{1}{\reynolds}\div\left\{ \eta(\grad\uv+\grad\uv^{\transp})
                                        \right\},
\]
$h_t + \div(h\uv) = 0$, $\div\uv =0$,
where we have define a scaled surfactant concentration $\phi=\Gamma/\mach^2$.
This model corresponds formally to the incompressible variable 
density Navier-Stokes 
equations with a density dependent viscosity, and with $\phi$ acting
as a pressure-like variable. The inviscid limit
$\reynolds\rightarrow\infty$ reduces to Ref.~\cite[Sec. 4.1.2.]{chomaz2001}.

We note that air drag effects
can significantly change soap film flow measurements
over flow distances of tens of centimeters\cite{rutgersAirDrag,huangAirDrag}. 
However, recent experiments operate in a regime
where air drag effects are secondary and the dominant forces
are due to the dynamics of the thin liquid layer\cite{wuSoapJFM2004}.
Future work will consider air drag 
effects\cite{rutgersAirDrag,huangAirDrag} in the 
model~\eqref{eq:mom}--\eqref{eq:soapConcentration}.

\soapSection{Small amplitude waves}
We discuss an ``acoustic'' limit as 
the simplest flow that can be used to illustrate compressibility 
effects.  Consider the small perturbations 
$\uv = 0 + \tilde{u}$, 
$\quad h =     1 + \tilde{h}$, 
and $\quad \Gamma = 1 + \tilde{\Gamma}$
on top of a quiescent base  flow $\bar{\uv}$,  $\bar{h}$, and $\bar{\Gamma}$.
The leading order linear equation is
\begin{equation}\label{eq:linearWave}
  \pder{^2 \tilde{u}}{t^2} = 
     -\frac{1}{\machBending^2}
        \pder{^4 \tilde{u}}{x^4} + \frac{1}{\mach^2}\,\pder{^2 \tilde{u}}{x^2}
     +\frac{4}{\reynolds}\,\pder{^2}{x^2}\pder{\tilde{u}}{t}.
\end{equation}
The last term in Eq.~\eqref{eq:linearWave} causes viscous dissipation
of the wave solutions and is omitted for simplicity in the following 
discussion.
The dispersion relation for solutions of the form 
$\tilde{u}=\exp(i(\omega t + kx))$ reveals  two types of limiting behavior.
Pure bending waves (for $\mach,\reynolds\rightarrow\infty$) are described by
\begin{eqnarray*}
\pder{^2 \tilde{u}}{t^2} &=& 
    -\frac{1}{\machBending^2} \pder{^4 \tilde{u} }{x^4}
    \qquad 
\end{eqnarray*}
and the dispersion relation  $ \omega = \pm k^2/\machBending$.
These waves are dispersive\cite{lighthill} with a group velocity 
$d\omega/dk =  2k/\machBending$ that is twice the phase velocity.
This is similar to the dispersive effect of the capillary terms
in water wave theory\cite{lighthill}.
The Navier-Stokes or Euler equations do not exhibit this
type of behavior.

Marangoni waves arise from the limit $\machBending,\reynolds\rightarrow\infty$,
where the small amplitude dynamics is governed by
\begin{eqnarray*}
  \pder{^2 \tilde{u}}{t^2} &=& 
     \frac{1}{\mach^2}\pder{^2 \tilde{u}}{x^2}.
\end{eqnarray*}
These elastic waves arise
from stretching of the film that induces tangential stresses to restore the 
equilibrium concentration. 
Marangoni waves are analogous to sound waves in gas dynamics with
the dispersion relation
$ \omega = \pm k/\mach$, and a (nondimensional)
phase and group velocity  $1/\mach$ that is independent of wavenumber. 
Slow flowing soap films
can be viewed as 2-d incompressible,
when the flow speed is a small fraction of the elastic sound 
speed ($\mach\ll 1$).
However, many experiments
operate in a regime where $\mach=\Ord(1)$ and 2-d compressibility
effects are clearly visible\cite{riveraVorobieffEcke}.
Marangoni waves have been measured in recent soap film
experiments\cite[Table 1]{wenSoapShock2}.

\soapSection{Derivation}
We present the details of the derivation 
for completeness and to simplify 
some aspects of past work\cite{idaMiksis2,chomaz2001}.
The dynamics of the soap film is governed by the three-dimension
incompressible Navier-Stokes equations with constant density $\rho$ 
and viscosity $\mu$. We define two-dimensional quantities
$\uv=(u,v)$, $\grad p=(p_x, p_y)$, and $\laplace u=u_{xx} + u_{yy}$
and use subscripts to denote derivatives.
We assume the soap film flow is symmetric about the 
center surface and impose $u_z =0$, $v_z =0$, $w=0$, $c_z=0$ at $z=0$.
Hence we ignore the bending mode and are only considering the 
symmetric (peristaltic) mode\cite{couderSoapFilm,chomaz2001}.
The thin layer of fluid is surrounded by free surfaces $z=\pm h(x,y,t)$ 
on which surfactant is transported. 
The surfactant concentration $c(x,y,z,t)$
in the interstitial fluid satisfies\cite{levichKrylov,chomaz2001}
an advection-diffusion equation with a diffusivity $D_c$ and
a free-surface boundary condition $D_c\partial c/\partial n= \jv$.
The flux $\jv= ( Kc - \Gamma)/\tau$ models
surfactant transport between the surface and the interior of the film
with transport parameters $K$, $\tau$.  %%\cite{chomaz2001}.
The use of both an interior and a surface concentrations $c$ and $\Gamma$
in a soap film model is due to Chomaz\cite{chomaz2001}.

The surfactant on the on the free surface $z=h(x,y,t)$ 
has a variable concentration $\Gamma$ that is transported 
according to~\cite{stoneSurfaceTransport}
\begin{equation}
\pder{\Gamma}{t} +\nabla_s\cdot\left(\Gamma\uv\right)
          + (2{\cal M}) (\nv\cdot\uv)\Gamma = D_{surf}\laplace_s\Gamma
          + \jv
\end{equation}
where 
the flow and derivative operators are restricted to the surface,
the flux $\jv$ allows exchange of surfactant with the interior 
concentration $c$, and  $2{\cal M}$ is
the mean curvature (see \cite{supplement,struik} for details).
We denote the average concentration $C_m$,
which yields the average surfactant concentration $\Gamma_m = KC_m$.
We scale $\lambda=L/(\tau U)$, and $\Kstar = K/H$.

The nondimensional bulk equations are, on dropping the primes,
\begin{eqnarray*}
 \matder{\uv} +w\pder{\uv}{z} 
    &=& ~ -\frac{1}{\mach^2}\grad p 
        + \frac{1}{\reynolds}\laplace \uv 
        + \frac{\eps^{-2}}{\reynolds}\pder{^2\uv}{z^2}\\
 \matder{w} + w\pder{w}{z}
    &=& ~-\frac{\eps^{-2}}{\mach^2}\pder{p}{z} 
        + \frac{1}{\reynolds}\laplace w 
        + \frac{\eps^{-2}}{\reynolds}\pder{^2 w}{z^2} \\
   \matder{c} +w\pder{c}{z} &=& ~~~\frac{1}{\schmidt\reynolds}\laplace c 
          +\frac{\eps^{-2}}{\schmidt\reynolds}\pder{^2 c}{z^2},
\end{eqnarray*}
and the divergence free condition $\div\uv + \partial w/\partial z =0$.

\soapSubsection{Boundary conditions on the free surface}

The free surface $z= h (x,y,t )$ satisfies a kinematic boundary condition
$ h_t + u h_x + v h_y = w$.
The viscous surface stresses are
balanced by capillary forces and Marangoni stresses arising
from the surfactant concentration distribution.
The nondimensional surface tension is given by
$   \sigma      = 1+ \zeta\left( 1- \Gamma \right),
$
where $\zeta = \sigma_r\Gamma_m/\sigma_m = \eps^2\reynolds/\mach^2$.
The scaling  $\zeta=\Ord(\eps^2)$
is determined by a distinguished limit arising in the tangential 
stress condition (see below).
The normal stress condition 
$\jump{ -p + 2\mu_0 \normalvec\cdot\rstrain\cdot\normalvec}
   = 2\sigma {\cal M}$ in nondimensional form is
\begin{eqnarray}\label{eq:normalStress}
\left( p + 2{\cal M}\left( 1+ \zeta(1-\Gamma)\right)\right)
&=& \frac{2\machBending^2}{\reynolds\xi^2}
      \left\{ h_xu_z + h_yv_z + \div\uv \right. \notag\\
 -\eps^2(h_x^2u_x + h_y^2 v_y &+&h_x h_y(u_y+v_x) \notag \\
            &-& h_xw_x - \left.h_y w_y)\right\}
            + \Ord(\eps^4),    \notag
\end{eqnarray}
where $\xi^2 = 1+ \eps^2|\grad h|^2$ and $2{\cal M}=\laplace h + \Ord(\eps^2)$\cite{supplement,struik}.

The tangential stress conditions 
$\jump{ \tangvec^1\cdot\rstrain\cdot \normalvec }
   = \tangvec^1\cdot\grad_s \sigma$ and
$\jump{ \tangvec^2\cdot\rstrain\cdot \normalvec }
   = \tangvec^2\cdot\grad_s \sigma$
in dimensionless form are
\begin{eqnarray*}
  u_z &+&\eps^2 [-2h_x u_x - h_y (u_y +v_x ) + w_x
                      -h_x^2 u_z  \\
      &-& h_x h_y v_z +2h_x^2 w_z]
=  -\eps^2\frac{\reynolds}{\mach^2} \Gamma_x + \Ord(\eps^4) \\
  v_z &+&\eps^2[ -2h_yv_y -h_x(u_y +v_x) +w_y
                     -h_y^2v_z     \\
      &-& h_x h_y u_z +2h_y^2w_z]
= -\eps^2\frac{\reynolds}{\mach^2} \Gamma_y + \Ord(\eps^4).
\end{eqnarray*}
Note that we have the choice of at least two scalings of the coefficient $\zeta$.
In the first case, $\zeta = \eps^2 \machBending^2/\mach^2=\Ord(\eps^2)$,
and the surfactant concentration dependent terms do not enter the dynamics
through the normal stress condition~\eqref{eq:normalStress} at leading order
which leads to the soap film 
model~\eqref{eq:mom}--\eqref{eq:soapConcentration}. 
This is the distinguished limit considered in Chomaz\cite{chomaz2001},
following earlier work~\cite{deWit,idaMiksis2}.
In the second case, $\zeta=\Ord(1)$,
the leading order fluid velocity would be slaved to the surfactant 
concentration\cite{deWit}. This is not a physically relevant scaling
for gravity driven fast soap film flows. The $\zeta=\Ord(1)$ scaling 
is  not considered here any further.

\soapSubsection{Asymptotic expansion}

We expand the solution 
$\qv=(h,u,v,\Gamma,c,p) =\bar{\qv} + \eps^2\tilde{\qv}+\Ord(\eps^4)$ in $\eps$.
We will obtain equations of motion depending only on $(x,y,t)$
by averaging across the thin layer in the $z$ direction.

%%\paragraph*{Leading order.}
The leading order tangential stress condition $\bar{\uv}_z=0$
implies $   \bar{\uv} = \bar{\uv}(x,y,t)$.
The vertical velocity $\bar{w}$ satisfies $\bar{w} = -z\div\bar{\uv}$.
The leading order velocity $\bar{\uv}=\bar{\uv}(x,y,t)$
is determined by the $\Ord(\eps^2)$ equations  as in
the derivation of thin jet models\cite{deWit,erneux}.
The normal stress condition~\eqref{eq:normalStress} at leading order
is
\[
    \bar{p} + \laplace\bar{h} = -\frac{2\machBending^2}{\reynolds}\,
           \left( \div\bar{\uv} \right).
\]
The leading order  mass conservation and surfactant transport equations are
$\bar{c}_z=0$,
\begin{eqnarray*}
  \bar{h}_t + \bar{u}\bar{h}_x + \bar{v}\bar{h}_y 
        &=& \bar{w} = -\bar{h}\div \bar{\uv}, \\
  \bar{\Gamma}_t + \bar{u}\bar{\Gamma}_x + \bar{v}\bar{\Gamma}_y 
       + \bar{\Gamma}(\div \uv)&=& \lambda(\bar{c} - \bar{\Gamma})
                  + \frac{1}{\schmidtGamma\reynolds}\laplace\bar{\Gamma}. 
\end{eqnarray*}
 
%\paragraph*{Order $\Ord(\eps^2)$.} 
The interstitial concentration satisfies 
$\tilde{c}_z=-\lambda\eps^2\reynolds\schmidt\Kstar(\bar{c}-\bar{\Gamma})$,
which allows integration of the leading order to obtain Eq.~\eqref{eq:soapConcentration}.
The momentum equations at $\Ord(\eps^2)$ are
\begin{eqnarray*}
   \frac{1}{\reynolds}\tilde{\uv}_z &=& \bar{h}\left(\matder{\bar{\uv}} 
             + \frac{1}{\machBending^2}\grad\bar{p} 
                                    -\frac{1}{\reynolds}\laplace\bar{\uv}\right),
\end{eqnarray*}
and the tangential stress conditions are
\begin{eqnarray*}
\tilde{u}_z &=& -\frac{\reynolds}{\mach^2}\Gammabar_x 
          + 2\bar{h}_x\ubar_x + \bar{h}_y(\ubar_y + \vbar_x)\\  
           &&\mbox{} + \bar{h}(\div\uvbar)_x  + 2\bar{h}_x\div\uvbar, \\
\tilde{v}_z &=& -\frac{\reynolds}{\mach^2}\Gammabar_y
          + 2\bar{h}_y\vbar_y + \bar{h}_x(\ubar_y + \vbar_x)  \\
           &&\mbox{} + \bar{h}(\div\uvbar)_y + 2\bar{h}_y\div\uvbar.
\end{eqnarray*}
The leading order momentum equation  is rewritten
in divergence form as
\begin{eqnarray*}
  \reynolds\left(\bar{h}\matder{\ubar}\right.  
             &+& \frac{1}{\mach^2} \bar{\Gamma}_x 
              - \left.\frac{\bar{h}}{\machBending^2}\laplace \bar{h}_x\right) \\
  &=&  2\derx\left( \bar{h} \ubar_x\right)
          + \dery\left( \bar{h} \ubar_y + \bar{h} \vbar_x \right) 
                 +2\derx\left( \bar{h}\div\uvbar \right), \\
  \reynolds\left(\bar{h}\matder{\vbar}\right. 
        &+& \frac{1}{\mach^2} \bar{\Gamma}_y
         -  \left.\frac{\bar{h}}{\machBending^2}\laplace \bar{h}_y \right)  \\
     &=&\derx\left( \bar{h} \vbar_x + \bar{h} \ubar_y \right) + 2\dery\left( \bar{h} \vbar_y \right) 
                      + 2\dery\left( \bar{h}\div\uvbar \right).
\end{eqnarray*}
Eq.~\eqref{eq:mom} follows since we can 
define the rate-of-strain tensor $\Dv = (\grad \uv + \grad \uv^{\transp})/2$
and use  $2\div\left(\bar{h}\Dv\right)$ to simplify the RHS.
This completes the derivation.

\soapSection{Discussion}

In summary, we have presented a quasi-two dimensional viscous compressible 
model of soap film flow that shares many similarities with 
the Navier-Stokes equations but that also presents
some intriguing differences.
Future research is needed
to determine the appropriate boundary conditions for the model
since the analysis herein and in past work\cite{chomaz2001} is only 
valid away from physical boundaries extending through the film.
Further, the vorticity $\omega = v_x - u_y$ in a soap film satisfies,
ignoring molecular diffusion ($\reynolds\rightarrow\infty$),
\begin{equation}\label{eq:vorticity}
\pder{\omega}{t} + \uv\cdot\grad\omega =  -\omega(\div\uv)
   + \frac{1}{h^2\,\mach^2}\,\grad h \cross \grad \Gamma.
\end{equation}
The baroclinic\cite{lesieur} term
$\grad h\cross\grad \Gamma = -h_y\Gamma_x +h_x\Gamma_y$
acts as  a vorticity source in regions where the isocontours
of thickness and surfactant concentration do not coincide,
This effect is absent from constant density Navier-Stokes flow.
Numerical  simulations using Eqs.~\eqref{eq:mom}--\eqref{eq:soapConcentration}
will be necessary to investigate 
the importance of baroclinic effects, surface 
bending elasticity contributions, viscosity
and compressibility in fast flowing soap film experiments.

\begin{acknowledgments}
We thank Mr. Pak-Wing Fok and Prof. Jonathan Goodman for discussions.
This work was performed under the auspices of the U.S. Department
of Energy by University of California Lawrence Livermore National
Laboratory under contract No. W-7405-Eng-48.
\end{acknowledgments}

%\bibliography{fastFlowReferences,epap}
%\bibliographystyle{unsrt}

%\makeLLNLBackCover
\newpage
\onecolumngrid 

\appendix
\begin{center}{\Large{\center{SUPPLEMENT:} \\
       A viscous compressible model of soap film flow 
       and its  \\
       equivalence with the Navier-Stokes equations 
       }}
\end{center}

\section{Soap film surface geometry}\label{app:surfaceGeometry}

In this section, we summarize differential geometry results
that are needed for describing the surface geometry of the
soap film. We derive expressions for the mean curvature
of the surface following Struik\cite[Sec. 2-7]{struik}.
We consider the dynamics of a thin fluid layer
that is symmetric about the center surface $z=0$ of the film.

The surface of the soap film is located at $z= \pm h(x,y,t)$.
In the following, the time dependence is implied and left out from the 
 notation.  All the expressions are dimensional. (Note that
in the non-dimensional, scaled, setting the surface height gradients
become $\grad h \rightarrow \eps\grad h$.)

We denote the surface vector by $\xivec(x,y) = (x,y, h(x,y))$.
The surface normal is
\begin{equation}
  \normalvec = \frac{\xivec_x \cross \xivec_y}{|\xivec_x \cross \xivec_y|}
   = \frac{ \left( -h_x, -h_y, 1 \right)}{\left( 1+ h_x^2 +h_y^2  \right)^{1/2}},
\end{equation}
where $\xivec_x=(1,0,h_x)$, $\xivec_y=(0,1,h_y)$, and
$|\xivec_x\cross \xivec_y|^2 = 1+h_x^2 +h_y^2$.
The surface tangent vectors are
\begin{equation}
\tangvec_1 = \frac{(1,0,h_x)}{\left( 1+ h_x^2 +h_y^2  \right)^{1/2}},
\quad
\tangvec_2 = \frac{(0,1,h_y)}{\left( 1+ h_x^2 +h_y^2  \right)^{1/2}}.
\end{equation}

The mean curvature is determined from the first and second fundamental form\cite{struik}
\begin{eqnarray}
I   &=& |\xivec_x|^2\,dx^2 + 2\xivec_x\cdot\xivec_y\,dx\,dy + |\xivec_y|^2\,dy^2, \\
II  &=& \xivec_{xx}\cdot\normalvec\,dx^2 
          + 2\xivec_{xy}\cdot\normalvec\,dx\,dy 
	  + \xivec_{yy}\cdot\normalvec\,dy^2,
\end{eqnarray}
where $ |\xivec_x|^2=1+h_x^2$, $|\xivec_y|^2 = 1+h_y^2$ and 
$2\xivec_x\cdot\xivec_y = h_x\,h_y$.
The fundamental forms simplify to
\begin{eqnarray*}
I  &=& (1+h_x^2)\,dx^2 + 2h_xh_y\,dx\,dx+(1+h_y^2)\,dy^2, \\
II &=& \frac{h_{xx}}{s}\,dx^2 + \frac{2h_{xy}}{s}\,dx\,dy + \frac{h_{yy}}{s}\,dy^2,
\end{eqnarray*}
where $s=(1+h_x^2 + h_y^2)^{1/2}$.
The mean curvature is, following Struik's notation\cite{struik},
\begin{eqnarray}
{\cal M} &=& \frac{1}{2}\,\frac{Eg - 2fF +eG}{EG-F^2} \\
  &=& \frac{1}{2}\,\frac{\laplace h + h_x^2h_yy -2h_xh_yh_{xy} +h_y^2 h_xx}
                        {\left( 1+ |\grad h|^2\right)^{3/2}}.
\end{eqnarray}
Here
\begin{alignat}{2}
E= 1+h_x^2, 
  &\quad
e= \frac{h_{xx}}{\left(1 + h_x^2 +h_y^2\right)^{1/2}} \\
F = h_x h_y, 
   &\quad
f= \frac{h_{xy}}{\left(1 + h_x^2 +h_y^2\right)^{1/2}}  \\
G = 1+h_y^2
   &\quad
g= \frac{h_{yy}}{\left(1 + h_x^2 +h_y^2\right)^{1/2}}.
\end{alignat}
and 
\begin{eqnarray*}
  EG-F^2 &=& (1+h_x^2)(1+h_y^2) - h_x^2 h_y^2 = s^2, \\
  Eg-2fF+eG &=& \left((1+h_x^2) h_{yy} -2h_xh_y h_{xy} +(1+h_y^2)h_{xx}\right)/s.
\end{eqnarray*}
For an almost flat surface we obtain $2{\cal M} \approx \laplace h $.

\section{Viscous surface stresses}

We summarize here the dimensional
viscous surface stresses at $\Ord(1)$ and $\Ord(\eps^2)$. 
Define $\xi^2 = 1+ |\eps\grad h|^2$.
First, recall the tangent vectors
$  \tangvec^1 = (1, 0, \eps h_x)/\xi $,
$  \tangvec^2 = (0, 1, \eps h_y)/\xi$,
and the normal vector $\normalvec = \tangvec^1 \cross \tangvec^2=
 ( -\eps h_x, -\eps h_y, 1 )/\xi$.
The viscous surface stresses are, to $\Ord(\eps^4)$,
\begin{eqnarray}
\normalvec\cdot\rstrain\cdot\normalvec
&=& \frac{U}{L\xi^2}\left(
   -h_x u_z - h_y v_z -\div\uv
   +\eps^2\left\{
       h_x^2 u_x + h_y^2 v_y +h_x h_y (u_y + v_x) -h_x w_x -h_y w_y
    \right\}
\right),  \\
\tangvec^1\cdot\rstrain\cdot\normalvec
&=& \frac{U}{2\eps  L \xi^2}\left(
u_z + \eps^2\left\{
      -2h_x u_x - h_y ( u_y + v_x) + w_x 
        - h_x^2 u_z - h_x h_y v_z + 2h_x^2 w_z
   \right\}
\right), \\
\tangvec^2\cdot\rstrain\cdot\normalvec
&=& \frac{U}{2\eps  L \xi^2}\left(
v_z + \eps^2\left\{
     -2h_y v_y -h_x (u_y + v_x) + w_y 
        - h_y^2 v_z - h_x h_y u_z + 2h_y^2 w_z 
   \right\}
\right).
\end{eqnarray}

\end{document}